\documentstyle[pra,aps]{revtex}
\begin{document}
\draft
\twocolumn[\hsize\textwidth\columnwidth\hsize\csname 
@twocolumnfalse\endcsname                            
\title{The Kohn mode for trapped Bose gases 
within  the dielectric formalism}
\author{J\"urgen Reidl$^1$, Gyula Bene$^2$, Robert Graham$^1$,
P\'eter Sz\'epfalusy$^{1,3}$}
\address{$^1${\it Fachbereich Physik, Universit\"at Gesamthochschule
Essen,
45117 Essen, Germany}\\
$^2${\it Institute for Theoretical Physics, E\"otv\"os University,
     P\'azm\'any P\'eter s\'et\'any 1/A, H-1117 Budapest, Hungary,}\\
$^3${\it Department of Physics of Complex Systems,
E\"otv\"os University, P\'azm\'any P\'eter s\'et\'any 1/A,
H-1117 Budapest, Hungary,\\
and
Research Institute for Solid State Physics and Optics,
P.O. Box 49, H-1525 Budapest, Hungary}}
\date{\today}
\maketitle

\begin{abstract}
The presence of undamped harmonic center of mass oscillations 
of a weakly interacting Bose gas in a harmonic trap is 
demonstrated within the dielectric formalism for a previously introduced 
finite temperature approximation including exchange. 
The consistency of the approximation with the Kohn theorem is thereby 
demonstrated. The Kohn modes are found explicitly, generalizing 
an earlier zero-temperature result found in the literature.
It is shown  how the Kohn mode disappears from the single-particle 
spectrum, while remaining in the density oscillation spectrum,
when the temperature increases from below to above the condensation 
temperature.
\end{abstract}


\vskip2pc]                                            
\narrowtext                                           

\section{Introduction}
In 1961 {\sl W. Kohn} \cite{kohn} first noticed that the electron-electron 
interaction does not change the cyclotron resonance frequency in 
a bulk three dimensional gas. {\sl Brey et al.} \cite{brey}
and {\sl Dobson} \cite{dobsen}
expanded the proof as a {\sl generalized Kohn theorem}
for parabolic quantum wells. Independent of the interactions the 
bare harmonic-oscillator frequency is preserved in the excitation 
spectra. 
These oscillations describe the center of mass movements along the 
main axes of the trap. They 
are used for calibration in the experiments with trapped 
Bose gases \cite{varenna}.\\
Many different model approximations were derived since the 
discovery of Bose condensation in trapped dilute Bose gases.
For them the satisfaction of the Kohn theorem provides an 
important consistency check.
That the Kohn theorem is satisfied 
within  the zero temperature Bogoliubov equations 
for  a trapped dilute Bose gas  
was proven by {\sl Fetter and Rokhsar} \cite{fetter}.
The Kohn theorem was demonstrated by {\sl Griffin and coworkers}
for various approximations leading to hydrodynamic equations
\cite{griffin}. {\sl Stoof and coworkers} checked the presence of the 
Kohn modes in an approximate variational approach to the kinetic theory 
also in the collisionless regime \cite{stoof}. 
A finite temperature theory based on a linear response formalism 
was given by {\sl Minguzzi and Tosi} \cite{Min} and the validity 
of the Kohn theorem in this approach was mentioned in their paper,
but without proof. \\
The purpose of the present paper is to examine the validity of the 
Kohn theorem in a specific finite temperature  approximation 
within the dielectric formalism.
It was introduced  and studied in a previous 
paper \cite{shifts}.
It was shown there that its results agree with those results
which can also be calculated in the {\sl Minguzzi and Tosi} approach
\cite{Min}, but that it gives additional results, in particular for 
the 1-particle Green's functions.  \\
The starting point is the  closed set of eigenvalue equations for 
the single particle Green's functions 
and the density-density correlation functions.
We first identify the eigenmodes for  the center of mass movement.
Then we show that they are eigenfunctions of the density 
autocorrelation function at the 
bare trap frequency. \\
We furthermore prove the validity of the Kohn theorem in 
the single particle spectra in the Bose condensed system
and show how the Kohn modes disappear from the single-particle 
spectrum, but remain in the density autocorrelation spectrum if the 
number of particles in the condensate goes to zero.

\section{Eigenvalue equations for the density  and single particle 
autocorrelation functions}
In this section we consider the eigenvalue equations  for the 
single particle excitations and  
the density autocorrelation function above and below 
the critical temperature  for Bose condensation $T_c$.
The eigenfrequencies $\omega_k$ and 
the eigenmodes $\delta n_{{\rm tot},k}
(\bbox{r})$ of the 
density correlation function 
$\chi (\bbox{r},\tau;\bbox{r}',\tau') =-
\langle T_\tau \left[
\tilde{n}(\bbox{r},\tau)\tilde{n}(\bbox{r}',\tau')\right]\rangle
$, 
where $\tilde{n}(\bbox{r})= \hat{n}(\bbox{r})-
\langle \hat{n}(\bbox{r})\rangle$,
are given by   
\begin{eqnarray}
\int d^3\,\bbox{r}'{\chi}^{-1}(\bbox{r},\bbox{r}',
\omega_k)\delta n_{{\rm tot},k}(\bbox{r}')=0\,.
\label{coll_modes}
\end{eqnarray}  
It is useful to express $\chi$ in terms of its interaction-irreducible 
part $\tilde{\chi}$ via
\begin{eqnarray}
&&{\chi}(\bbox{r},\bbox{r}',\omega)
\nonumber\\
&&\quad=\tilde{\chi}(\bbox{r},\bbox{r}',\omega)+\frac{g}{\hbar}\int\,d^3\bbox{r}_1\,
\tilde{\chi}(\bbox{r},\bbox{r}_1,\omega)
{\chi}(\bbox{r}_1,\bbox{r}',\omega)\,.
\label{eq_chi}
\end{eqnarray}  
Here and in the following the {\sl tilde}-sign denotes 
{\sl interaction-irreducible}
quantities, which are given by diagrams which cannot be split into 
two by cutting a single interaction line.\\ 
Then Eq. (\ref{coll_modes}) takes the form
\begin{eqnarray}
\delta n_{{\rm tot},k}(\bbox{r})=\frac{g}{\hbar}
\int d^3\,\bbox{r}'\tilde{\chi}(\bbox{r},\bbox{r}',
\omega_k)\delta n_{{\rm tot},k}(\bbox{r}')\,.
\label{dysonchi}
\end{eqnarray}
Besides we consider the matrix of Green's 
functions
\begin{equation}
G_{\alpha,\beta}(\bbox{r},\tau;\bbox{r}',\tau')=-
\langle T_\tau\left[ \hat\varphi_\alpha(\bbox{r},\tau)
\hat\varphi_\beta^\dagger(\bbox{r}',\tau')\right]\rangle
\label{eq:gfdef}
\end{equation}
with  field operators in Matsubara representation $\hat\varphi_1
(\bbox{r},\tau) \equiv \hat\varphi(\bbox{r},\tau)$ and
$\hat\varphi_2 (\bbox{r},\tau) \equiv
\hat\varphi^\dagger (\bbox{r},\tau)$.
The corresponding eigenfrequencies $\omega_k'$ and eigenfunctions
$\varphi_{\alpha,k}$ are determined by
\begin{eqnarray}
\int d^3 \bbox{r}' G^{-1}_{\alpha\beta}(\bbox{r},\bbox{r}',\omega_k')
\varphi_{k,\beta}(\bbox{r}')
=0\,.
\label{greeneigen}
\end{eqnarray} 
Of course we need to 
make approximations for $\tilde{\chi}$ and $G_{\alpha\beta}$
and describe now the 
model we previously introduced within the 
framework of the dielectric formalism \cite{shifts,hom}.
The approximations will be further detailed  in 
the subsections \ref{suba} and \ref{subb} and the appendix
where they  will be used  to derive  the eigenvalue equations
for the  different excitation spectra.
At this point it is sufficient to fix the approximations 
for the thermal density $n_{\rm T}(\bbox{r})$. The 
external trap potential $U(\bbox{r})$ is
\begin{equation}
U(\bbox{r})=\frac{1}{2}m\sum_{i=1,2,3}\Omega_i^2 x_i^2\,,
\end{equation}
where $\Omega_i$ are the main trap frequencies.\\
The effective interaction potential  $v(\bbox{r},\bbox{r}')$ has the form 
$v(\bbox{r},\bbox{r}')=g
\delta(\bbox{r}-\bbox{r}')$ with the constant interaction strength 
$g$.
It is taken to contain already the complete series of ladder-diagrams 
describing repeated two-particle scattering, i.e. it is the 2-particle
T-matrix.\\
The stationary thermal density $n_{\rm T}(\bbox{r})$ is calculated 
using  as auxiliary quantity 
the stationary Hartree-Fock Greek's function $G^{HF}$
\begin{eqnarray}
\left(\hbar\omega-\hat{H}^{HF}(\bbox{r})\right)
G^{HF}(\bbox{r},\bbox{r}',\omega)&=&\hbar\delta(\bbox{r}-\bbox{r}')\,.
\end{eqnarray}
The Hartree-Fock Hamiltonian is 
\begin{equation}
\hat{H}^{HF}(\bbox{r})
=-{\hbar^2 \over 2m}\Delta+U(\bbox{r})-\mu 
+2g n_{\rm tot}(\bbox{r})\,,
\label{Hartree-Fock}
\end{equation}   
with the chemical potential $\mu$ and the total density 
$n_{\rm tot}(\bbox{r})=n_{\rm T}(\bbox{r})+
n_{\rm c}(\bbox{r})$, where  $n_{\rm c}$ is the condensate density and
$n_{\rm T}$ is 
\begin{eqnarray}
n_{\rm T}(\bbox{r})&=&\frac{k_B T}{\hbar}
\lim_{\eta\to 0}
\sum_n e^{i\omega_n\eta}G^{HF}(\bbox{r},\bbox{r},\omega_n)
\nonumber\\&
=&\sum_k n(\varepsilon_k) |\varphi^{HF}_k(\bbox{r})|^2\,.
\label{ntdet}
\end{eqnarray}
$\omega_n$ are the Matsubara frequencies
$2n\pi k_B T/\hbar$ and 
$n(\varepsilon_k)$ is the Bose factor to the eigenenergy 
$\varepsilon_k$ for  the eigenfunction $\varphi^{HF}_k(\bbox{r})$  
of  the Hartree-Fock Hamiltonian. The latter will be taken as real,
in the following, without restriction of generality because the 
Hamiltonian (\ref{Hartree-Fock}) is real.
Within the  model \cite{shifts} all internal 
lines of diagrams are calculated with $G^{HF}$.

\subsection{The uncondensed Bose gas\label{suba}}
We first consider the uncondensed Bose gas.
Here, 
the lowest order contribution  to $\tilde{\chi}$ is  
the bubble diagram $\tilde{\chi}^0$
containing two Hartree-Fock Green's functions
\begin{eqnarray} 
&&\tilde{\chi}^0(\bbox{r},\bbox{r}',\omega)  =
\nonumber\\
&&-{1 \over \beta\hbar}
\sum_n G^{HF}(\bbox{r},\bbox{r}',i\omega_n)
G^{HF}(\bbox{r}',\bbox{r},i\omega_n-\omega)\\
&&=
\sum_{k,l}\varphi^{HF}_k(\bbox{r})\varphi^{HF}_l(\bbox{r})
\frac{\left(n(\varepsilon_k)-n(\varepsilon_l)
\right)}{\omega+(\varepsilon_k-\varepsilon_l)/\hbar}
\varphi^{HF}_k(\bbox{r}')\varphi^{HF}_l(\bbox{r}')\nonumber
\\
&&\overbrace{=}^{l\leftrightarrow k}
\frac{1}{2}
\sum_{k,l}\varphi^{HF}_k(\bbox{r})\varphi^{HF}_l(\bbox{r})\left(n(\varepsilon_k)-n(\varepsilon_l)
\right)\nonumber\\&&
\times\left[\frac{1}{\omega+(\varepsilon_k-\varepsilon_l)/\hbar}
\nonumber\right.\\&&\,\,\,\left.
+\frac{1}{-\omega+(\varepsilon_k-\varepsilon_l)/\hbar}\right]
\varphi^{HF}_k(\bbox{r}')\varphi^{HF}_l(\bbox{r}')\,.
\label{chi0}
\end{eqnarray}
Here we first performed  the Matsubara sum 
and then symmetrized the obtained expression
with respect to $k\leftrightarrow l$.\\
$\tilde{\chi}$ is represented, within the model, 
by summing up  the  multiple particle-hole scattering processes
including exchange with the result, in symbolic notation
\begin{eqnarray}
&&\tilde{\chi}(\bbox{r},\bbox{r}',\omega)=
\nonumber\\
&&\quad\int\,d^3\bbox{r}_1\,\left[1-\frac{g}{\hbar}
\tilde{\chi}^0(\omega)\right]^{-1}(\bbox{r},\bbox{r}_1)
\tilde{\chi}^0(\bbox{r}_1,\bbox{r}',\omega)\,.
\label{eq:chir}
\end{eqnarray}  
Inserting  these diagrams $\tilde{\chi}$ in Eq. (\ref{dysonchi}),
multiplying from the left with the integral 
operator $\left[1-(g/\hbar)
\tilde{\chi}^0(\omega)\right]$ and combining similar terms, 
the density correlation spectra are given by 
\begin{eqnarray}
\delta n_{{\rm tot},k}(\bbox{r})=
2\frac{g}{\hbar}\int d^3\,\bbox{r}'\tilde{\chi}^0(\bbox{r},\bbox{r}',
\omega_k)\delta n_{{\rm tot},k}(\bbox{r}')\,.
\label{thermfluc2}
\end{eqnarray}  
Above $T_c$ the single particle Green's function in our model 
is just the Hartree-Fock approximation. 
Of course the corresponding eigenfrequencies 
$\omega_k'=\varepsilon_k/\hbar$ given by
\begin{eqnarray}
\int d^3 \bbox{r}'
\{G^{HF}\}^{-1}(\bbox{r},\bbox{r}',\omega_k')
\varphi_k^{HF}(\bbox{r}')=
\varphi_k^{HF}(\bbox{r})\,
\end{eqnarray}
differ from those of the density correlation spectra $\omega_k$.
\subsection{The Bose condensed gas\label{subb}}
In the presence of the condensate the two kind of excitations 
(\ref{coll_modes}) and (\ref{greeneigen}) share
the same frequencies. This fact is a consequence of the underlying
spontaneous symmetry breaking and can be conveniently formulated
by using the dielectric formalism. This formalism expresses all the
relevant quantities in terms of proper irreducible quantities (whose
graphs remain connected even if one cuts an arbitrary, single
interaction line or propagator). 
For the Bose condensed gas  
we have to include interaction  processes with the condensate 
atoms, involving many further diagrams within 
our approximation. 
In order to avoid a somewhat lengthy detour 
we refer the interested 
reader to the appendix, where we briefly summarize the approximations
and the derivation of the following equations.
These approximations have been presented in detail in reference \cite{hom}.
The complete set of equations for the 
eigenfunctions within our model are 
(Eq.(\ref{eigeq3}))
\begin{eqnarray}
\left(\tilde{H}_0(\bbox{r})+gn_{\rm c}(\bbox{r})\right)\varphi_{1,k}(\bbox{r})
+g \Phi^2_0(\bbox{r})\varphi_{2,k}(\bbox{r})=&&\nonumber\\
\hbar\omega_k\varphi_{1,k}(\bbox{r})-
2g\Phi_0(\bbox{r})\delta n_{{\rm T},k}(\bbox{r})&&\,,
\label{ueq}\\
\left(\tilde{H}_0(\bbox{r})+gn_{\rm c}(\bbox{r})\right)\varphi_{2,k}(\bbox{r})
+g\Phi^{*2}_0(\bbox{r})\varphi_{1,k}(\bbox{r})=&&
\nonumber\\-\hbar\omega_k\varphi_{2,k}(\bbox{r})-
2g\Phi_0^*(\bbox{r})\delta n_{{\rm T},k}(\bbox{r})&&\,,
\label{veq}
\end{eqnarray}
where we defined
\begin{eqnarray}
\delta n_{{\rm c},k}(\bbox{r})=
\Phi_0^*(\bbox{r})\varphi_{1,k}(\bbox{r})+\Phi_0(\bbox{r})
\varphi_{2,k}(\bbox{r})
\label{ncdef}\,
\end{eqnarray}
and 
\begin{eqnarray}
\delta n_{{\rm T},k}(\bbox{r})=
\delta n_{{\rm tot},k}(\bbox{r})-\delta n_{{\rm c},k}(\bbox{r})\,.
\label{ntdef}
\end{eqnarray}
Due to the presence of the condensate we have quasiparticle  
excitations with two-component  vector fields 
$\varphi_{k,\alpha}(\bbox{r})$ ($\alpha=1,2$) as 
eigenfunctions.\\
The  condensate wave function 
$\Phi_0(\bbox{r})$ is calculated from  
the finite-temperature 
generalization of the Gross-Pitaevskii-equation
\begin{equation}
\tilde{H}_0(\bbox{r})\Phi_0(\bbox{r})=0\,,
\label{eq:gpeq}
\end{equation}
with the Hamiltonian $\tilde{H}_0(\bbox{r})$
\begin{equation}
\tilde{H}_0(\bbox{r})=-{\hbar^2 \over 2m}\Delta+U(\bbox{r})-\mu +
g|\Phi_0(\bbox{r})|^2
+2gn_{\rm T}(\bbox{r})\,.
\label{eq:gp}
\end{equation}
We allow complex solutions of Eq. (\ref{eq:gp}), i.e. our considerations
apply also to the case where vortices are present.\\
The total density is given by  the sum $n_{\rm T}+n_{\rm c}$ with 
(\ref{ntdet}) and 
$n_{\rm c}(\bbox{r})=|\Phi_0(\bbox{r})|^2$.\\
In the appendix we show (see Eq. (\ref{dnt2})) 
that in the presence of a Bose condensate 
Eq. (\ref{thermfluc2}) must be rewritten as 
\begin{eqnarray}
\delta n_{{\rm T},k}(\bbox{r})=
2\frac{g}{\hbar}\int d^3\,\bbox{r}'\tilde{\chi}^0(\bbox{r},\bbox{r}',
\omega_k)\delta n_{{\rm tot},k}(\bbox{r}')\,,
\label{thermfluc}
\end{eqnarray}
which can be also expressed in terms of $\delta n_{{\rm c},k}$
\begin{eqnarray}
&&\delta n_{{\rm T},k}(\bbox{r})=
2\frac{g}{\hbar}\int d^3\,\bbox{r}_1\int d^3 \,\bbox{r}_2\nonumber\\&&
\times\left[
1-2\frac{g}{\hbar}
\tilde{\chi}^0(\omega_k)
\right]^{-1}(\bbox{r},\bbox{r}_1)
\tilde{\chi}^0(\bbox{r}_1,\bbox{r}_2,
\omega_k)
\delta n_{{\rm c},k}(\bbox{r}_2)\,.
\label{thermfluc3}
\end{eqnarray}
With Eq. (\ref{ncdef}) we see that Eq. (\ref{ueq}) and (\ref{veq}) 
still form a set of homogeneous coupled equations 
for $\varphi_{\alpha,k}$.\\
For $n_{\rm c}=0$ 
Eq. (\ref{ueq}) agrees with the Hartree-Fock equation.
In the low temperature region  
$T\to 0$ we have $\tilde{\chi}^0\to 0$ and according 
to Eq.(\ref{thermfluc}) $\delta n_T\to 0$. Therefore, at $T\approx0$ 
Eqs. (\ref{ueq}) and (\ref{veq}) approach the usual Hartree-Fock 
Bogoliubov Popov equations.\\
Eqs. (\ref{ueq}) and (\ref{veq}) agree with the  
linearization of the time dependent Gross-Pitaevskii equation 
around its stationary solution $\Phi_0(\bbox{r})$ if the  
thermal density is assumed to be  time-dependent and is 
also linearized around its stationary value $n_{\rm T}(\bbox{r})$.
Of course we then need additional equations like
Eqs. (\ref{thermfluc3}) and (\ref{ncdef}) to fix $\delta n_{{\rm T},k}$. 
\section{The Kohn Theorem}
The Kohn theorem states that the center of mass oscillations of the 
total density $n_{\rm tot}$ along the main axes 
of the trap are unchanged by the 
interactions and appear at the bare 
trap frequencies $\Omega_i$ as exact eigenvalues.\\
For the proof  we choose arbitrarily one of those directions $\bbox{e_i}$.
The  center of mass movement is just a 
displacement of the 
total density in the trap. \\
Therefore, the eigenfunctions to the Kohn frequency $\Omega_i$ should be 
given by the   
infinitesimal displacement operations 
$\eta\,\partial_i$ acting 
on the stationary densities 
($\eta\ll 1$) 
$\eta\,\partial_i n(\bbox{r})= n(\bbox{r}+\eta\,\bbox{e}_i)
-n(\bbox{r})=\delta n_i(\bbox{r}) $.\\
The Kohn mode $\delta n_{{\rm tot},i}(\bbox{r})$ 
should therefore be proportional to 
$\partial_i n_{\rm tot}(\bbox{r})
=\partial_i n_{\rm c}(\bbox{r})+2 \sum_k n(\varepsilon_k) 
\varphi^{HF}_k(\bbox{r})
\partial_i \varphi^{HF}_k(\bbox{r})$ if we insert the expression 
for $n_{\rm T}$.\\
We prove that the Kohn theorem is satisfied 
in two steps where we set $\omega=\Omega_i$
in  all the  expressions.
First we show that we get $\delta n_{{\rm T},i}=\partial_i n_{\rm T}$
on the left hand side of Eq. (\ref{thermfluc}) if we insert  
$\delta n_{{\rm tot},i}=\partial_i n_{\rm tot}$ on the right hand side.
Using the result
$\delta n_{{\rm T},i}=\partial_i n_{\rm T}$ in  
Eqs. (\ref{ueq}) and (\ref{veq}) we calculate 
the corresponding $\varphi_{\alpha,k}$ from which 
we derive $\delta n_{{\rm c},i}=\partial_i n_{\rm c}$.
We hereby reobtain $\delta n_{{\rm tot},i}=\partial_i n_{\rm tot}$
proving that $\partial_i n_{\rm tot}$ and $\varphi_{\alpha,k}$ 
are eigenfunctions of the equations to the trap frequency $\Omega_i$.
In the following 
the effect of the partial differential operator on the 
eigenfunctions of $\hat{H}^{HF}$ are 
evaluated  by using   the commutator relations 
\begin{eqnarray}
\left[\partial_i,\hat{H}^{HF}(\bbox{r})\right]&=&
m\Omega_i^2x_i+2g\partial_i n_{\rm tot}(\bbox{r})\,,\\
\left[x_i,\hat{H}^{HF}(\bbox{r})\right]&=&
\frac{\hbar^2}{m}\partial_i\,.
\end{eqnarray}
We obtain the 
matrix elements of the coordinate and the 
differential operators within the basis of  the Hartree-Fock eigenfunctions
\begin{eqnarray}
(\varepsilon_k-\varepsilon_l)
\langle  l|x_i|k\rangle&=&\frac{\hbar^2}{m}\langle  l|\partial_i|k\rangle\,,
\label{identb}\\
(\varepsilon_k - \varepsilon_l )
\langle l|\partial_i|k\rangle&=&m\Omega_i^2\langle l|x_i|k\rangle\nonumber\\&&
+2g\langle l|\partial_i n_{\rm tot}
(\bbox{r})|k\rangle \label{identa}\,,
\end{eqnarray}
where we used the abbreviation
\begin{equation}
\langle l|\hat{A}|k\rangle= \int\,d^3\bbox{r}\varphi^{HF}_l(\bbox{r}) \hat{A}
\varphi^{HF}_k(\bbox{r})\,.
\end{equation}
In the first step we have to insert 
$\delta n_{{\rm tot},k}=\partial_i n_{\rm tot}$ 
in the  expression (\ref{chi0}) for  the right hand side 
of Eq. (\ref{thermfluc}) with $\omega=\Omega_i$
\begin{eqnarray}
\delta n_{{\rm T},i}(\bbox{r})
&=&g\sum_{k,l}\varphi^{HF}_k(\bbox{r})\varphi^{HF}_l(\bbox{r})
\left(n(\varepsilon_k)-n(\varepsilon_l)
\right)\nonumber\\&\times&
\left[\frac{\langle k|\delta n_{{\rm tot},i}|l\rangle}
{\hbar\Omega_i+(\varepsilon_k-\varepsilon_l)}
+\frac{\langle k|\delta n_{{\rm tot},i}|l\rangle}
{-\hbar\Omega_i+(\varepsilon_k-\varepsilon_l)}\right]\,.
\label{seceq}
\end{eqnarray}
We further concentrate on the coefficients corresponding to 
$\varphi^{HF}_k(\bbox{r})\varphi^{HF}_l(\bbox{r})
\left(n(\varepsilon_k)-n(\varepsilon_l)
\right)$
\begin{eqnarray}
&&g\left[\frac{\langle l|\delta n_{{\rm tot},i}|k\rangle}
{\hbar\Omega_i+(\varepsilon_k-\varepsilon_l)}
+\frac{\langle l|\delta n_{{\rm tot},i}|k\rangle}
{-\hbar\Omega_i+(\varepsilon_k-\varepsilon_l)}\right]
\nonumber\\
&&=2g\left[
\frac{
(\varepsilon_k-\varepsilon_l)\langle l|\delta n_{{\rm tot},i}|k\rangle}
{(\varepsilon_k-\varepsilon_l)^2-(\hbar\Omega_i)^2}
\right]\\
&&\overbrace{=}^{(\ref{identa})}
\left[
\frac{(\varepsilon_k-\varepsilon_l)^2 \langle l|\partial_i |k\rangle
-(\varepsilon_k-\varepsilon_l)m\Omega_i^2\langle l|x_i |k\rangle }
{(\varepsilon_k-\varepsilon_l)^2-(\hbar\Omega_i)^2}
\right] \\
&&\overbrace{=}^{(\ref{identb})}
\left[
\frac{(\varepsilon_k-\varepsilon_l)^2 \langle l|\partial_i |k\rangle
-(\hbar\Omega_i)^2\langle l|\partial_i |k\rangle }
{(\varepsilon_k-\varepsilon_l)^2-(\hbar\Omega_i)^2}
\right] \\
&&=\langle l|\partial_i|k\rangle\,.
\end{eqnarray}
We obtain the result
\begin{eqnarray}
&&\delta n_{{\rm T},i}(\bbox{r})\nonumber\\&
&=\sum_{k,l}\langle l|\partial_i|k\rangle\varphi^{HF}_k(\bbox{r})\varphi^{HF}_l(\bbox{r})
\left(n(\varepsilon_k)-n(\varepsilon_l)
\right)\\
&&=2\sum_{k,l}n(\varepsilon_k)\varphi^{HF}_k(\bbox{r})\varphi^{HF}_l(\bbox{r})
\langle l|\partial_i|k\rangle\,,
\label{dntres}
\end{eqnarray}
where in the last step we used the antisymmetry of \\
$\langle l|\partial_i|k\rangle$ 
under exchange of $k$ and $l$. 
The expression (\ref{dntres}) for $\delta n_{{\rm T},i}$ is 
exactly the partial differential of $n_{\rm T}$ with respect to 
$x_i$. Therefore, we get by comparing the left and the right hand 
side of Eq. (\ref{thermfluc}) $\delta n_{{\rm T},i}=
\partial_i n_{\rm T}$.\\
In the second step we have to insert $\delta n_{{\rm T},i}=
\partial_i n_{\rm T}$ in Eqs. (\ref{ueq}) and (\ref{veq}).
But then it is straight forward to see that  Eqs. (\ref{ueq}) and (\ref{veq})
are solved for $\omega=\Omega_i$ and $\delta n_{{\rm T},i}=
\partial_i n_{\rm T}$
by setting
\begin{eqnarray}
\label{kohn1}
\varphi_{1,i}(\bbox{r})&=&(\partial_i-x_i m\Omega_i/\hbar )\Phi_0(\bbox{r})
\,,\\
\varphi_{2,i}(\bbox{r})&=&(\partial_i+x_im\Omega_i/\hbar )
\Phi_0^*(\bbox{r})\,.
\label{kohn2}
\end{eqnarray}
Inserting these expressions in Eq. (\ref{ncdef}) we get 
$\delta n_{\rm c}(\bbox{r})=\partial_i n_{\rm c}(\bbox{r})$ 
and hence
$
\delta n_{\rm tot}(\bbox{r})=\delta n_{\rm c}(\bbox{r})+
\delta n_{\rm T}(\bbox{r})$\\
$=\partial_i n_{\rm tot}(\bbox{r})$, completing the proof.
Therefore the expressions $\partial_i n_{\rm tot}(\bbox{r})$ 
and $\varphi_{\alpha,i}(\bbox{r})$ are eigenfunctions 
with the main trap frequencies $\Omega_i$ as eigenfrequencies. \\
{\sl Fetter and Rokhsar} \cite{fetter} proved  
that (\ref{kohn1}), (\ref{kohn2}) are the Kohn modes of the
Bogoliubov equations at $T=0$. Our preceding considerations 
generalize their proof to finite $T$.
In the whole calculation we have chosen to fix the normalization 
of the 
density fluctuations $\delta n_{{\rm tot},i}=\partial_i n_{\rm tot}$.
By performing a simple partial integration we get the 
normalization of the Kohn eigenfunctions 
$\varphi_{1,i}(\bbox{r}),\varphi_{2,i}(\bbox{r})$ according to 
$\int d^3 \bbox{r}(|\varphi_{1,i}(\bbox{r})|^2-|\varphi_{2,i}(\bbox{r})|^2)
=2m\Omega_i N_{\rm c}/\hbar$,
where $N_{\rm c}$ is the total number of atoms in the condensate.
With $N_{\rm c} \to 0$ the Kohn modes are seen to disappear from the 
single-particle spectrum.\\
For $N_{\rm c}=0$ (i.e. $\Phi_0=0=n_c$) 
Eq. (\ref{ueq}) agrees with  the usual 
Hartree-Fock equation where the 
Kohn mode is no longer present in the single particle spectra.
But the Kohn modes survive  in the density correlation spectra
given by   Eq. (\ref{thermfluc}) 
for $n_{\rm tot}=n_{\rm T}$. 
In this case the  proof of the Kohn theorem 
is already finished after the first step of our proof.\\

\section{Discussion and Summary}
The fundamental difference between a spatially homogeneous and a trapped 
Bose gas is the absence of the conservation las of momentum in the latter,
which plays an important role in the theory of homogeneous systems. 
Certainly any approximation made in homogeneous systems should satisfy 
this fundamental law. In the special case of harmonic trapping potentials
the momentum conservation law is replaced by Kohn's theorem, which states
that there are three exact special modes where the center of mass 
oscillates harmonically with the three main trap frequencies. 
While this is easily proven for the exact Hamiltonian, the question 
whether any given approximation still satisfies this theorem is, in general,
a rather nontrivial one. In the case of Bose condensed gases the question 
acquires an additional aspect, because of the coincidence of single particle
and density oscillation spectra in Bose condensates. 
The Kohn mode, which is a density oscillation mode, must then have 
a single particle counterpart, which somehow disappears from the spectrum
if the temperature is raised, so that the condensate disappears
and the single particle spectrum and the density oscillation spectrum 
become decoupled.\\
In the present paper we have examined these questions for a specific
approximation which includes direct interaction and exchange and is 
formulated within the dielectric formalism, which guarantees form the 
onset the coincidence of single particle and density oscillation 
spectra in the Bose condensed regime. By deriving a closed set 
of equations (\ref{ueq})-(\ref{ntdef}) for the single particle 
modes and the density modes generalizing the usual Bogoliubov-deGennes  
equations in Popov approximation, and solving them for 
the special coupled modes, corresponding for the densities just to a 
translation, we could not only verify the Kohn theorem for the closed 
set of equations, but obtain explicit expressions for the 
single particle modes (\ref{kohn1}), (\ref{kohn2}). These 
expressions for the latter are beautifully simple. They generalize
the corresponding zero-temperature result obtained by 
{\sl Fetter and Rokhsar} by replacing the zero-temperature solution 
of the Gross-Pitaevskii equation by its finite counterpart
of our specific model within the dielectric formalism.
In this form our results (\ref{kohn1}), (\ref{kohn2}) show explicitly 
how the single particle component of the Kohn mode vanishes 
while keeping the density oscillations component unchanged, if 
the number of particles in the condensate is sent to 
zero.\\
A simpler version of the model treated here leaves out all 
exchange processes. This is then the simplest model of an 
interacting Bose gas which one can set up within the dielectric formalism.
It was analyzed in detail 
for a homogeneous Bose gas in 
\cite{kondor}, but one can also analyze it for the trapped 
gas along the lines of \cite{shifts} extending work in \cite{bene}.
Using the same procedure as employed in the present paper one can 
readily show that the Kohn theorem is respected also by this 
simpler model.

\appendix
\section*{}
In the first part of the appendix we 
introduce the necessary quantities given by our model 
\cite{shifts,hom} including exchange
to describe the density autocorrelation function for $T<T_c$.
Then we show how Eqs. (\ref{ueq}),(\ref{veq}), which we have 
used here as a convenient starting point, are obtained.
We first note that the contributions to $\tilde{\chi}$ above $T_c$ given 
in Eq.(\ref{eq:chir}) are not only {\sl interaction line irreducible}
but also {\sl propagator line irreducible}. 
We denote these so-called {\sl regular} contributions by $\tilde{\chi}^{(r)}$.
Below $T_c$   
we get  
contributions to $\tilde{\chi}$ in addition to the regular ones, namely   
\cite{shifts}
\begin{eqnarray}
\tilde{\chi}(\bbox{r},\bbox{r}',\omega)&=&
\tilde{\chi}^{(r)}(\bbox{r},\bbox{r}',\omega)
+\int d^3 \bbox{r}_1\int d^3\bbox{r}_2
\tilde{\Lambda}^{(r)}_\alpha(\bbox{r},\bbox{r}_1,\omega)
\nonumber\\&&
\times
\tilde{G}_{\alpha\beta}(\bbox{r}_1,\bbox{r}_2,\omega)
\tilde{\Lambda}^{(r)*}_\beta(\bbox{r}_2,\bbox{r}',\omega)\,.
\label{eq:chis}
\end{eqnarray}
The interaction line irreducible Green's functions 
$\tilde{G}_{\alpha\beta}$ are related to  the exact 
Green's functions $G_{\alpha\beta}$ by the Dyson-type equation
\begin{eqnarray}
&&G_{\alpha,\beta}(\bbox{r},\bbox{r}',\omega)=
\tilde{G}_{\alpha,\beta}(\bbox{r},\bbox{r}',\omega)
\nonumber\\
&&+\frac{g}{\hbar}
\int\,d^3\bbox{r}_1\,\int\,d^3\bbox{r}_2\,\int\,d^3\bbox{r}_3\,
\tilde{G}_{\alpha,\gamma}(\bbox{r},\bbox{r}_1,\omega)\nonumber\\
&&\times \tilde{\Lambda}^*_{\gamma} (\bbox{r}_1,\bbox{r}_2,\omega)
\Lambda_\delta (\bbox{r}_2,\bbox{r}_3,\omega)
G_{\delta,\beta}(\bbox{r}_3,\bbox{r}',\omega)\,,
\label{eq_G}
\end{eqnarray} 
where the vertex functions $\Lambda_{\alpha}$ are related 
to their regular parts by 
\begin{eqnarray}
&&\Lambda_{\alpha}(\bbox{r},\bbox{r}',\omega)
=\tilde \Lambda_{\alpha}(\bbox{r},\bbox{r}',\omega)\nonumber\\
&&+
\frac{g}{\hbar}\int\,d^3\bbox{r}_1\,
\tilde{\chi}^{(r)}(\bbox{r},\bbox{r}_1,\omega)
\Lambda_\alpha(\bbox{r}_1,\bbox{r}',\omega)\,.
\label{eq_lambda}
\end{eqnarray} 
The regular  vertex functions 
$\tilde{\Lambda}^{(r)}_\alpha$ express 
the coupling of the single-particle Green's functions
$\tilde{G}_{\alpha\beta}$ to the density autocorrelation 
function $\tilde{\chi}$.
The coupling comes  from  excitation processes out of 
and relaxation processes into  the 
condensate.\\
In the model introduced in \cite{shifts,hom} the 
vertex functions $\tilde{\Lambda}^{(r)}_\alpha$ contain 
the lowest order contributions given by $\Phi_{\alpha,0}^*(\bbox{r})$ 
and a series of higher order diagrams 
corresponding to the diagrams summed up in Eq. (\ref{eq:chir})
\begin{eqnarray}
\tilde{\Lambda}^{(r)}_\alpha (\bbox{r},\bbox{r}',\omega)
&=&\Phi^*_{\alpha,0}(\bbox{r})\delta (\bbox{r}-\bbox{r}')
+\frac{g}{\hbar}
\int d^3 \bbox{r}_1\tilde{\Lambda}^{(r)}_\alpha (\bbox{r},\bbox{r}_1,\omega)
\nonumber\\&&\times
\tilde{\chi}^0(\bbox{r}_1,\bbox{r}',\omega)\,. 
\label{eq:lambdar}
\end{eqnarray}
If we allow $\Phi_0$ to be complex we have to 
use $\Phi_0$ for the excitation processes due to 
the corresponding annihilation of a condensate atom and 
$\Phi_0^*$ in the opposite case.
In order to simplify the notations 
we write  $\Phi_{\alpha,0}$ ($\alpha=1,2$) with 
$\Phi_{1,0}=\Phi_0$ and $\Phi_{2,0}=\Phi_0^*$. \\
$\tilde{G}_{\alpha\beta}$ is in general expressed by the regular part
$\tilde{\Sigma}^{(r)}_{\alpha\beta}$ of the self-energies in the form 
\begin{eqnarray}
&&\left[\frac{1}{\hbar}
\tilde{G}(\omega)\right]^{-1}(\bbox{r},\bbox{r}')
=\delta (\bbox{r}-\bbox{r}')\nonumber\\&&
\,\times\left(\begin{array}{cc}
\hbar\omega+\frac{\nabla^2\hbar^2}{2m}+\mu-U(\bbox{r})& 0\\
0 & -\hbar\omega+\frac{\nabla^2\hbar^2}{2m}+\mu-U(\bbox{r}) 
\end{array}\right)
\nonumber \\&& \,
-\tilde{\Sigma}^{(r)}(\bbox{r},\bbox{r}',\omega)\,.
\end{eqnarray}
The approximation of the model for $\tilde{\Sigma}^{(r)}$ is defined 
by 
\begin{eqnarray}
\tilde{\Sigma}^{(r)}_{\alpha \beta}(\bbox{r},\bbox{r}',\omega)&=&
\delta (\bbox{r}-\bbox{r}')(g n_{\rm c}(\bbox{r}')+2gn_{\rm T}(\bbox{r}'))
\delta_{\alpha\beta}\,,\nonumber
\\&&  
+\frac{g^2}{\hbar}
\Phi_{\alpha,0}(\bbox{r})\tilde{\chi}^{(r)}(\bbox{r},\bbox{r}',\omega)
\Phi_{\beta,0}^*(\bbox{r}')\nonumber\,.
\\&&  
\end{eqnarray}  
Like in Eqs. (\ref{eq:chir}),(\ref{eq:lambdar}) we add to the 
lowest order  regular self-energy diagrams 
$g n_{\rm c}(\bbox{r}')+2gn_{\rm T}(\bbox{r}')$ appearing in the diagonal 
of 
$\tilde{\Sigma}^{(r)}$
the sum over multiple particle-hole scattering processes including 
exchange  
with  additional factors $\Phi_0^*$, $\Phi_0$ due to excitation processes 
out of and the absorption processes into the condensate.\\ 
Below $T_c$ the eigenfunctions $\varphi_1(\bbox{r})$ and 
$\varphi_2(\bbox{r})$   
can be expressed as a linear functional 
of the corresponding eigenfunction $\delta n_{{\rm tot},k}(\bbox{r})$
of the density autocorrelation function spectra
({\sl see} \cite{bene})
\begin{eqnarray}
\varphi_{\alpha,k}(\bbox{r})&=&\int d^3 \bbox{r}_1 \int d^3 \bbox{r}_2
\sum_{\beta}\tilde{G}_{\alpha\beta}(\bbox{r},\bbox{r}_1,\omega_k)
\nonumber\\&&\times\tilde{\Lambda}^{(r)*}(\bbox{r}_1,\bbox{r}_2,\omega_k)
\delta n_{{\rm tot},k}(\bbox{r}_2)\,.
\label{compeig}
\end{eqnarray}
Without any approximations for the regular 
quantities $\tilde{\Sigma}^{(r)}_{\alpha\beta}$, $\tilde{\chi}^{(r)}$ and 
$\tilde{\Lambda}^{(r)}_\alpha$ the functions 
$\varphi_\alpha(\bbox{r})$ would fulfill Eq. (\ref{greeneigen}) 
if $\delta n_{{\rm tot},k}(\bbox{r})$ fulfills Eq. (\ref{dysonchi}).\\
We insert  Eq. (\ref{eq:chis}) in Eq. (\ref{dysonchi}) and take into account
(\ref{compeig})
to get
\begin{eqnarray}
\delta n_{{\rm tot},k}(\bbox{r})&=&\int d^3\bbox{r}'\left[\frac{g}{\hbar}
\tilde{\chi}^{(r)}(\bbox{r},\bbox{r}',\omega_k)
\delta n_{{\rm tot},k}(\bbox{r}')
\right.\nonumber\\&&
\left.+\tilde{\Lambda}^{(r)}_\alpha(\bbox{r},\bbox{r}',\omega_k)
\varphi_{\alpha,k}(\bbox{r}')\right]
\end{eqnarray}
Defining 
$\delta n_{{\rm c},k}(\bbox{r})=\Phi_0^*(\bbox{r})\varphi_{1,k}(\bbox{r})
+\Phi_0(\bbox{r})\varphi_{2,k}(\bbox{r})$
and separating   out the 
the lowest order 
diagram $\Phi^*_{\alpha,0}$ of 
$\tilde{\Lambda}^{(r)}_\alpha$  we derive an 
equation for 
$\delta n_{{\rm T},k}=\delta n_{{\rm tot},k}-\delta n_{{\rm c},k}$
\begin{eqnarray}
\delta n_{{\rm T},k}(\bbox{r})
&=&\frac{g}{\hbar}\int d^3 \bbox{r}'
\tilde{\chi}^{(r)}(\bbox{r},\bbox{r}',\omega_k) 
\nonumber\\
&&\times
\left(
\delta n_{{\rm c},k}(\bbox{r}')+\delta n_{{\rm tot},k}(\bbox{r}')
\right)
\label{ntsimp}\,.
\end{eqnarray}
Eliminating $\delta n_{{\rm c},k}$ by using 
$\delta n_{{\rm c},k}=\delta n_{{\rm tot},k}-\delta n_{{\rm T},k}$
this equation is solved by 
\begin{eqnarray}\delta n_{{\rm T},k}(\bbox{r})
=2\frac{g}{\hbar}\int d^3 \bbox{r}'
\tilde{\chi}^{0}(\bbox{r},\bbox{r}',\omega_k)
\delta n_{{\rm tot},k}(\bbox{r}')\,, 
\label{dnt2}
\end{eqnarray}
as can be verified by using the result (\ref{eq:chir}) for 
$\tilde{\chi}^{(r)}$.\\
Acting with  the operator   $\left[\frac{1}{\hbar}
\tilde{G}(\omega_k)\right]^{-1}_{\alpha\beta}(\bbox{r},\bbox{r}')
$ on ${\varphi}_{\beta,k}(\bbox{r}')$
and using (\ref{eq:chis})-(\ref{compeig}) we obtain
\begin{eqnarray}
&&(\pm\hbar\omega_k-\tilde{H}_0(\bbox{r}))\varphi_{
{1\atop 2},k}(\bbox{r})\nonumber\\&&
=\frac{g^2}{\hbar}
\int d^3 \bbox{r}'\Phi_{{1\atop 2},0}(\bbox{r})
\tilde{\chi}^{(r)}(\bbox{r},\bbox{r}',\omega_k)
\delta n_{{\rm c},k}(\bbox{r}')\nonumber\\&&
+g\Phi_{{1\atop 2},0}(\bbox{r})
\int d^3 \bbox{r}'\left(\delta(\bbox{r}-\bbox{r}')
+\frac{g}{\hbar}\tilde{\chi}^{(r)}(\bbox{r},\bbox{r}',\omega_k)\right)
\nonumber\\&&
\times\delta n_{{\rm tot},k}(\bbox{r}')\,,
\label{eigeq1}
\end{eqnarray}
where $\tilde{H}_0$ is given by (\ref{eq:gp}).
Eqs. (\ref{eigeq1}) can be simplified 
with the help of Eq. (\ref{ntsimp}) and we can directly give the results
\begin{eqnarray}
&&(\pm\hbar\omega_k-\tilde{H}_0(\bbox{r}))\varphi_{
{1\atop 2},k}(\bbox{r})
=\nonumber\\&&\quad
g\Phi_{{1\atop 2},0}(\bbox{r})\left(
\delta n_{{\rm tot},k}(\bbox{r})+\delta n_{{\rm T},k}(\bbox{r})\right)
\label{eigeq3}\,.
\end{eqnarray}
Eqs. (\ref{eigeq3}) agree with 
Eqs. (\ref{ueq}) and (\ref{veq}) using the definition  
$\delta n_{{\rm c},k}=\Phi_0^*\varphi_{1,k}+\Phi_0\varphi_{2,k}$.
\vskip0.5cm 
{\bf  Acknowledgment}
\vskip0.5cm 
The research results were attained with the assistance of the Humboldt  
Research Award to one of us (P.Sz.) and through support by a project of the  
Deutsche Forschungsgemeinschaft and the Hungarian Academy of Sciences under  
grant No 130. Support by the Deutsche Forschungsgemeinschaft
within its Sonderforschungsbereich 237 'Unordnung und grosse Fluktuationen',
and by the Hungarian National Research Foundation under Grant Nos. \mbox{T 029752},
\mbox{T 029552} and by the Hungarian Academy of Science under 
Grant No. AKP 9820$\,$2.2
is also gratefully acknowledged. 
One of us (Gy.B.) would like to thank the Hungarian
Academy of Sciences for support as a J\'anos Bolyai fellow.

\end{document}